\documentstyle[prl,aps,multicol,epsf]{revtex}
\begin{document}
\draft
\title{Scaling in a temperature quench in systems with a
Lifshitz point: Nonconserved and conserved order parameters}
\author{
Abhik Basu$^{1,2}$ and Jayanta K Bhattacharjee$^3$}
\address
{$^1$Poornaprajna Institute of Scientific Research, Bangalore, India,\\
$^2$Abteilung Theorie, Hahn-Meitner-Institut, Glienicker Strasse
  100, D-14109 Berlin, Germany,\\
$^3$ Department of Theoretical Physics, Indian Association for the
Cultivation of Science, Calcutta 700032, India.}
\maketitle
\sloppy
\begin{abstract}
We study the growth of an $N$-component (including $N=1$)
order parameter when a system with a Lifshitz point is quenched
from the homogeneous disordered state to the ordered states.
We study the scaling behaviours of the structure factors for both the
non-conserved and conserved order parameters in the long time
limit after a quench through the Lifshitz point 
by using the large-$N$ and renormalisation group (RG) methods. 
We construct the
analogues of Allen-Cahn and Lifshitz-Slyzov growth laws for nonconserved and 
conserved
order parameters which agree with our RG results for $N=1$. By extending our 
large-$N$
methods to the anisotropic Lifshitz point we show that the anisotropy is 
relevant for
the growth of the nonconserved order parameter, but irrelevant for the
 conserved case in the large-$N$ limit. 
We discuss the effects of mode coupling terms on scaling of the structure 
factor for the conserved order parameter case. We also consider the ordering
dynamics after a quench through an off-Lifshitz point. In a large-$N$ set up we
calculate the form of the structure factors of a non-conserved order
parameter when a system is quenched from the homogeneous
paramegnetic to the modulated phase. We show that after a
quench from the paramegnetic to the modulated phase the form of the structure
factor violates the standard form of dynamical scaling. Our results compare
favourably with the available numerical results.
\end{abstract}

\begin{multicols}{2}
\section{Introduction}
The problem of `phase-ordering kinetics' is the growth of order following
a temperature quench from the high temperature disordered state to the low
temperature state (for a general review see \cite{bray}). These 
systems serve as examples of typical non-equilibrium systems.
The structure factor exhibits temporal scaling after
a quench from the high temperature phase to the low temperature phase of an
Ising order parameter. The dynamics 
depends upon whether the order parameter is conserved or not. Following
the success of the large-$N$ type calculation in the critical phenomena
(see, for example \cite{ma}), the large-$N$ method has been extended 
to the problem of quench also \cite{bray}. 
In the long time limit,
this method yields same dynamic exponent as the renormalisation group (RG)
arguments, when the
dynamics is non-conserved, but when the dynamics is conserved, it yields two
time scales - one corresponding to the dynamic exponent as given by the RG
arguments, and a second one, differing from the first one by a factor
of $\log(t)$ where $t$ is the time elapsed after quench. However, subsequently, 
numerical
results for finite $N$ as well as analytical calculations in a $1/N$ expansion
for large-$N$ show that this temporal multiscaling is an essentially an
artefact of the $N\rightarrow \infty$ limit \cite{bray,braypap}. For finite $N$
one recovers simple scaling.

In this paper we consider the ordering dynamics of non-conserved (NCOP) and conserved
(COP) order parameters
after a temperature quench in a system with a Lifshitz point 
by using large-$N$ and
RG methods. We obtain several interesting results: (i) The relevance of the anisotropy
for the ordering dynamics of NCOP but its irrelevance in COP for large-$N$, (ii) RG
analyses for arbitrary $N$ for both NCOP and COP for the isotropic Lifshitz point. 
Our $N=1$ RG results agree with our analogues of Allen-Cahn equation
 for NCOP and
Lifshitz-Slyzov result for COP, (iii) The relevance of mode-coupling terms (torque)
for COP with $N=3$ and (iv) the form of the structure factor after a  quench
through an off-Lifshitz point (line EF; see Fig.(\ref{fig1})) in a large-$N$
approximation from the paramagnetic
phase to the modulated phase: We find that the standard form for the structure
factor exhibiting dynamical scaling breaks down. 
The organisation of the paper is as follows: In Sec.II we set up the
equations of motion and discuss the general scaling form for the structure 
factor and the
two-time correlator. In Sec.IIIA we calculate the dynamic exponent $z$ 
and the two-time exponent $\lambda$ after a quench through the Lifshitz point
for the NCOP by
using large-$N$ as well as RG methods. We discuss our analogue of the Allen-Cahn
result for $N=1$. We also elucidate the relevance of anisotropy on the ordering dynamics
in a large-$N$ set up. In Sec.IIIB we calculate $z$ 
for the COP after a quench through the Lifshitz point by using large-$N$ and RG
methods for the isotropic case. We set up our analogue of the 
Lifshitz-Slyzov law 
heuristically and show that it matches with our RG result for $N=1$. 
We then show that
the anisotropy is irrelevant for the COP in the large-$N$ limit. 
We also discuss the
relevance of mode-coupling terms in the equations of motion of COP in a 
simple set up. In Sec.IV, we use the large-$N$ method to calculate the 
structure factor after a quench from the paramagnetic 
to the modulated phase.  In Sec.V we conclude.

\section{Equation of motion for our system}
\begin{figure}[h]
\epsfxsize=13cm
\centerline{\epsffile{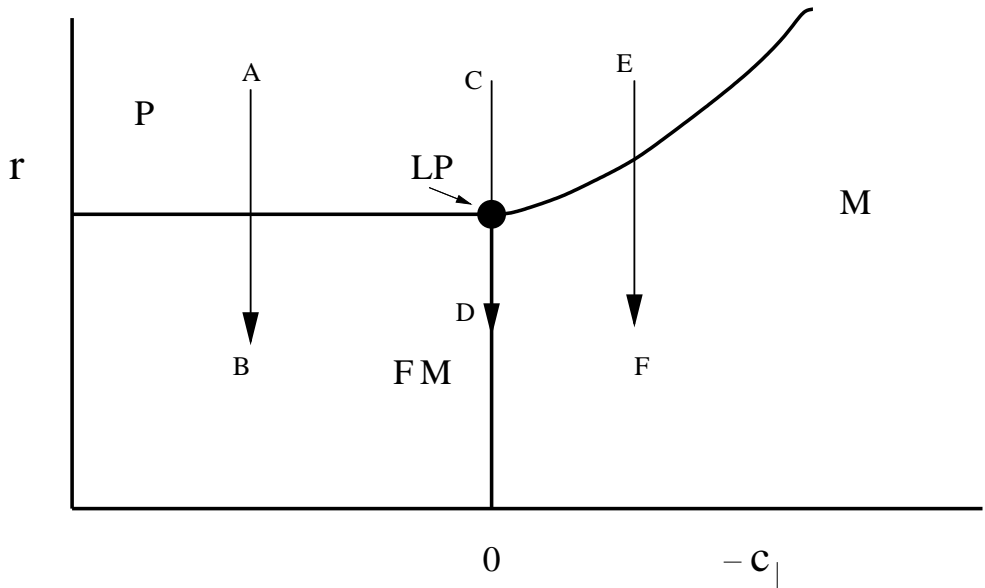}}
\caption{A schematic phase diagram for the Landau free energy Eq.(\ref{free})
for a vector order parameter. Here the high-temprature paramagnetic phase (P),
the low-temperature ferromagnetic phase (FM) and the low-temperature modulated
phase (M) meet at the Lifshitz point (LP). Arrows AB, CD and EF refer to
quench-paths. The phase boundaries and quench paths are schematic only.}
\label{fig1}
\end{figure}

The Landau-Ginzburg free energy for a system with $(d-m)$ Lifshitz point is 
\cite{horn,chaikin}
\begin{eqnarray}
F&=&{1\over\ 2}\int d^dx[r\phi^2+c_{||}(\nabla_{||}\phi)^2+c_{\perp}
(\nabla_{\perp}\phi)^2 \nonumber \\&+&D(\nabla^2 \phi)^2]+u\int d^dx (\phi^2)^2,
\phi^2=\phi_{\nu}\phi_{\nu};\nu=1,...,N,
\label{free}
\end{eqnarray}
where $\bf x=(x_{\perp},x_{||})$ is divided  into $m$ perpendicular
and $(d-m)$ parallel components.  ($d-m$) Lifshitz point is known as the $m$-axial
Lifshitz point; the special case for $m=d$, i.e., ($d-d$) Lifshitz point is known
as isotropic Lifshitz point. The field $\phi$ can be a scalar or an
$N$-component vector ($N=1$ corresponds to the scalar field case). 
When both $c_{||}$ and $c_{\perp}$ are positive, the
ordered phase is spatially uniform, but when $c_{\perp} <0$, the system
can lower energy by creating spatially modulated structures with wavevectors
$\mid k\mid=\mid c_{\perp}\mid /2D$. The point $r=0, c_{\perp}=0$ is a Lifshitz
point. There are three phases in the phase diagram: The paramagnetic
disordered phase, the low temperature ferromagnetic phase and the modulated
phase. A schematic phase diagram for a vector order parameter of a system with
a Lifshitz point is shown in Fig.(\ref{fig1}). 
An example of Lifshitz point is the system of mixtures of homopolymers
and diblock polymers \cite{pol}. It also arises in metamagnets with appropriate
choice of exchanges. A well-known model is the Anisotropic Next Nearest
Neighbour-interaction Ising (ANNNI) model.
There have been a great deal of work on the static critical proprties at the
Lifshitz point; recent references include Frisch, Kimball and Binder 
~\cite{binder}, Diehl, Shpot and Zia ~\cite{diehl},
Shpot and Diehl ~\cite{shpot}, Leite ~\cite{leite}, Pleimling and
Henkel ~\cite{pleimung}. There have been few studies on the dynamical properties
at the Lifshitz point; see, e.g., Huber ~\cite{huber}, Folk and Selke ~\cite{folk},
Selke ~\cite{selke}, and Selke ~\cite{domb}. Growth of order after
quenches in the ANNNI model has been studied by Kaski, Ala-Nissil\"{a} and Gunton
\cite{gunton}.
Our results are complementary to these works. Here,
we study quench along the
lines CD ($c_{\perp}=0$ and $r$ changes sign from positive to negative) and EF
($c_{\perp}$ changes sign from positive to negative).
For the $(d-d)$ Lifshitz point, all $c_{\mid\mid}$ are zero. 
\end{multicols}

\begin{multicols}{2}
The equation of motion for NCOP is
${\partial \phi\over\partial t}=-\Gamma {\delta F\over\delta \phi}$
where as, for COP, $\Gamma$ above 
is to be replaced by $\Gamma \nabla^2$.
Since there is no time translation invariance of the system, the two-time
correlation function $C({\bf k},t,t')=\langle \phi_{\bf k}(t)\phi_{-\bf
k}(t')\rangle$ will be a function of $t$ and $t'$
separately, unlike e.g. in dynamic critical phenomena where it is a function
of $t-t'$.  
In the asymptotically long-time limit $C({\bf k},t,t')$ typically 
exhibits a scaling form
\begin{eqnarray}
C({\bf k},t,t')&=&L(t)^d(L(t')/L(t))^{\lambda}g(kL(t))\nonumber \\
&=&t^{d/z}(t'/t)^{\lambda/z}
g(kL(t)),
\label{stanc}
\end{eqnarray}
where $L(t)\sim t^{1/z}$ is the characteristic length scale of the system at 
time $t$, $z$ is the dynamic exponent, $\lambda$ is the two-time exponent
and $f(0)$ is a constant. This scaling
form is expected to hold when both $t,t'>>t_o$ where $t_o$ is some microscopic
timescale and $L(t)\gg L(t')$. 
The structure factor or the equaltime correlation function is given by
$S(k,t)\equiv \langle\phi({\bf k},t)\phi(-{\bf k},t)\rangle$.
In the long time limit, $S(k,t)$ is expected to exhibit a scaling form
\begin{equation}
S(k,t)\sim t^{d/z}g(kt^{1/z}),
\label{stanform}
\end{equation}
 where $z$ is the dynamic exponent. We calculate 
$z$ and $\lambda$ in the long time limit given the equations of motion above.
In the large-$N$ limit, the scalar field $\phi$
should be replaced by $\phi_{\nu}$ with $\nu=1,..,N$. Expressions (\ref{stanc})
and (\ref{stanform}) have been shown to hold good after quenches where the
initial (high temperature) phase is random and the final (low temperature) 
phase is uniform. However, in a system with a Lifshitz point the following
possibilities arise: a)A quench from the paramagnetic to the ferromagnetic
phase through an off-Lifshitz point (along the line AB in Fig.(\ref{fig1}): 
The ordering dynamics here is same as 
that of an Ising order parameter
which has been discussed in details in the literature, b)A quench from the
paramagnetic to the ferromagnetic through the
 Lifshitz point (along the line CD),
and c)A quench from the
paramagnetic phase to the modulated phase (along the line EF).
Expressions (\ref{stanc}) and
(\ref{stanform}) are expected to hold for case b); however it is not apriori
clear if they hold true for case c) also. We investigate that here.

\section{Ordering dynamics after a quench through the Lifshitz point}
\subsection{Nonconserved order parameter}
We first calculate the structure factor in the long time limit after a quench
through the Lifshitz point (along the line CD in Fig.(\ref{fig1}) in the
large-$N$ approximation. In this limit, the equation of motion for the
$(d-d)$ Lifshitz point becomes (after scaling away $r$)
\begin{equation}
{\partial \phi_{\alpha}\over\partial t}=-\nabla^4 \phi_{\alpha}+\phi_{\alpha}
-{u\over\ N}(\phi_{\nu}\phi_{\nu})\phi_{\alpha}.
\label{eqsol}
\end{equation}
The initial conditions are given as
$\langle \phi_i({\bf k},0)\rangle =0, 
\langle \phi_i({\bf k},0)\phi_j({\bf -k},0)\rangle = \Delta\delta_{ij},$
where $\bf k$ is a wavevector.
We solve Eq.(\ref{eqsol}) in a self-consistent one-loop approximation 
which is equivalent to approximate Eq.(\ref{eqsol}) by (exact in the $N\rightarrow
\infty$ limit)
\begin{eqnarray}
{\partial \phi\over\partial t}=-\nabla^4 \phi+a(t)\phi,
\label{nclarge}
\end{eqnarray}
This amounts to replace
$\phi_{\nu}\phi_{\nu}/N$ by its average $<\phi^2>$ in the equation of motion.
Here $\phi$ stands for any of the $N$ components of $\phi_{\alpha}$. This
equation can also be derived in a diagrammatic method \cite{newman}.
Here $a(t)$ has to be solved selfconsistently. The formal solution of
Eq.(\ref{nclarge}) is given by
\begin{equation}
\phi_k (t)=\phi_k(0)\exp[-k^4 t+b(t)],
\end{equation}
where $b(t)=\int_0^t dt' a(t')$ implying
$a(t)={db\over\ dt}=1-\Delta\sum_k \exp[-2k^4t+2b(t)].$
The prescribed initial condition has been used to eliminate $\phi_k(0)$.
We claim that $a(t)$ is negligible in the large $t$ limit (which we will
justify {\em a posteriori}). Hence, on putting $a=db/dt=0$ we get
\begin{eqnarray}
1=\Delta\int d^dk \exp [2b-2tk^4] 
=C \exp(2b) t^{-d/4},
\end{eqnarray}
where $C$ is a constant.
We scale out the $t$ dependence by using a scaled variable $y=tk^4$ and 
absorbed numerical factors including $\Delta$ within $t$. Hence,
$b= (d/8) \ln t$, which immediately gives
\begin{equation}
S(k,t)=t^{d/4}\exp (-2k^4 t),
\end{equation}
which is of the conjectured scaling form. It is easy to see that the
{\em dynamic exponent} $z$=4. Now with the obtained value of $b$,
$a={db\over dt}\sim {1\over t} \rightarrow 0 \,\,\,{\rm as}\,\,\,
t\rightarrow \infty,$
which is the {\em a posteriori} justification of the assumption
$a(t)\rightarrow 0$ as $t\rightarrow \infty$. Our analysis immediately tells us that in
the long time limit,
\begin{equation}
\phi_{\bf k}(t)=\phi_{\bf k}(0)t^{d/8}\exp[-k^4 t].
\end{equation}  
This gives, for the unequal time correlation function
\begin{equation}
C(k,t,t')=\langle\phi_{\bf k}(t)\phi_{\bf -k}(t')\rangle=t^{d/4}\left({t'\over t}\right)
^{d/8}\exp[-k^4(t+t')],
\end{equation}
thus giving $\lambda=d/2$ same as for an Ising order parameter in the
$N\rightarrow \infty$ limit \cite{bray}. It will be very interesting to see
whether $\lambda$ for this system remains the same as that an Ising order
parameter even after inclusions of $O(1/N)$ corrections.

For NCOP, 
however $z$ cannot be calculated exactly (unlike 
COP; see below) through RG arguments,
but these arguments lead to the existence of another length scale
in the problem \cite{bray,bray1}.
In the Langevin equation for NCOP we add a white
noise representing thermal noise.
\begin{equation}
{\partial \phi_{\alpha}\over\partial t}=
-\Gamma {\delta F\over\delta \phi_{\alpha}}+\zeta_{\alpha},
\end{equation}
with $\langle\zeta_{\alpha}({\bf k_1},t_1)\zeta_{\nu}({\bf k_2},t_2)\rangle
=2\Gamma T\delta_{\alpha,\nu}
\delta (t_1-t_2)\delta({\bf k_1+k_2})$ such that the canonical distribution be
recovered in the equilibrium. Let under rescaling $k=k'/b, t'=b^z t$
\begin{equation}
\phi_{\alpha}(k,t)=\phi_{\alpha}(k'/b, b^zt')=b^{\chi}\phi'_{\alpha}(k',t').
\end{equation}
Hence the structure factor  scales as $S(k,t)=
b^{2\chi-d}t^{d/z}g(kt^{1/z})$. The choice of $\chi=d/2$ ensures that
$<\phi'(k',t')\phi'(-k',t')>\equiv S'(k',t')=S(k,t).$
At the $T=0$ fixed point, the Hamiltonian is not scale invariant, but has a
non-zero scaling dimension $y$: $F(\{{b^{\chi}\phi}\})=b^yF({\phi})$. After
rescaling the Langevin equation reads
\begin{equation}
b^{2\chi-y-z}{1\over\Gamma'}{\partial {\phi'}_{\alpha}(k',t')\over\partial t'}
=-{\delta F\over\delta {\phi'}_{\alpha}}=\zeta_{\alpha}'(k',t')/{\Gamma'}
\label{coarse}
\end{equation}
where the new noise $\zeta_{\alpha}'(k',t')=b^{\chi -y}\zeta_{\alpha}(k,t)$ and
the noise correlator $\langle \zeta_{\alpha}(k',t')'\zeta_{\nu}(-k',0)\rangle =
b^{d-z-2y}2T\Gamma\delta (t')$.
Thus we find, using $\chi=d/2$,
\begin{eqnarray}
{1/\Gamma'}&=&b^{d-y-z}h(b)(1/\Gamma),\\
T'&=&b^{-y}T.
\end{eqnarray}
where $h(b)$ is the contribution due to the elimination of the small
scales. The scale dependence of temperature $T$ appears essentially as the
coarse-grained Langevin equation has been rewritten in the form (\ref{coarse}).
The Hamiltonian is dominated by the $(\nabla ^2\phi)^2$ term.
One can construct 
the Porod law \cite{bray} for this problem which tells us
\begin{equation}
S({\bf k},t)\sim {1\over L^N k^{d+N}}.
\end{equation}
The energy density $\epsilon$ is given by
\begin{eqnarray}
\epsilon &\sim& \langle \mid \nabla^2 \phi\mid ^2\rangle\nonumber \\
&=&\int k^4 \langle \phi_{\bf k}\phi_{\bf -k}\rangle\nonumber \\
&=&\int d^dk {k^4\over L^N k^{d+N}}\nonumber \\
&\Rightarrow& \epsilon
\sim L^{-4} \,\,\,{\rm for}\,\,\, N>4 ;\;\nonumber \\
&\epsilon&\sim L^{-N}\zeta^{4-N} \,\,\,{\rm for}\,\,\, N<4.
\end{eqnarray}
where $\zeta=1/k_{max}$ is the UV cutoff. It is obvious that for large
$N$, $y=d-4$. For $N<4$, the dominant contribution comes from $L$ (which is the
lower cut off or the system size)
Hence, for $N=1$, $y=d-1$; 
$N=2$, $y=d-2$ and for $N=3$, $y=d-3$. For $N=4$, $y=d-4$ and in addition to
that there is an additional logarithm factor.

Thus for $N=1$ we substitute $y=d-1$ and obtain
${1\over\Gamma'}=b^{1-z} h(b){1\over\Gamma}.$
Under coarse graining in the free theory $\Gamma$ change only under rescaling: since
$z=4$ in the free theory $\Gamma'\sim b^3 \Gamma$. The nonlinear term becomes important
when $L(t)\gg w$ the interface thickness; 
in that regime $\Gamma$ reaches its fixed
point value $\Gamma^*$ which can be determined by matching with the 
value in the free theory
(when $L(t)\lesssim w$) \cite{bray}. 
This gives $\Gamma^*=w^3\Gamma$ ($\omega$ is measured in
units of the lattice spacing). We now argue this (which in turn suggests $z=4$)
is consistent with the
more conventional dynamical scaling at the critical point $T=T_c$. At $T<T_c$
the domain length $L(t)\sim (\Sigma \omega^3\Gamma t/M^2)^{1/z}$ where $\Sigma$
is the surface tension and $M$ is the magnetisation (see \cite{bray} for the
corresponding Ising case). Thus for $z=4$ we have 
$L(t)\sim (\Sigma \omega^3\Gamma t/M^2)^{1/4}$. Here $\Sigma$, $\Gamma$ etc.
are functions of $T$. In the critical region,
$\Sigma\sim\zeta^{-(d-1)},\,M^2\sim \zeta^{-(d-4+\eta)},\,\Gamma\sim
\zeta^{4-\eta-z_c}$ where $\zeta$ is the correlation length, $\eta$ is the
anomalous dimension and $z_c$ is the critical dynamical exponent of NCOP. Also
the interfacial thickness $\omega\sim \zeta$.  Substituting we find
\begin{equation}
L(t)\sim \zeta (t/\zeta^{z_c})^{1/4},
\label{scalscale}
\end{equation}
which is consistent with the conventional dynamical scaling \cite{bray}. 
This provides justification of $z=4$ and the existence of another length scale
$\omega$ in the problem.
This can also be seen in an Allen-Cahn type approach for 
this model (similar to Ref.\cite{bray} for an NCOP in the Ising model): We
get for a flat equilibrium profile ${d^4\phi\over d g^4}=V'(\phi)$ where $g$ is the
coordinate normal to the interface. Also, noting ${\partial\phi\over\partial t}=-
{\partial \phi\over\partial g}{\partial g\over\partial t}$ and by using the equation for
the equilibrium profile, we find ${\partial \phi\over\partial g}=\nabla^2\nabla\cdot 
{\bf \hat g}$ 
where we have neglected terms containing higher order derivatives of $\phi$ with
respect to $g$ for gently curving walls. Here ${\bf \hat g}$ is a unit vector normal to
the wall (in the direction for increasing $\phi$). Since ${\partial \phi\over\partial
g}$ is just the wall velocity $v$ we get $v=-\nabla^2 \nabla\cdot {\bf \hat g}$ which is
the analogue of the Allen-Cahn equation for this model. 
$\nabla^2 \nabla\cdot {\bf \hat g}$ is the
Laplacian of the mean curvature. Since for a single characteristic length $L$,
 velocity
$v\sim {dL\over dt}$ and $\nabla^2 \nabla\cdot {\bf \hat g}\sim {1\over L^3}$ giving 
$L(t) \sim t^{1/4}$ for non-conserved scalar order parameter. So we find $z=4$ same as
predicted by the RG analysis.

For $N=2, 3$, i.e., for vector order parameters, the Hamiltonian renormalises as
a free theory but eventually flows to the strong-coupling fixed point characterised by
$\phi^2=const.$, which eliminates one degree of freedom from the problem but the
remaining ones still renormalise as free fields suggesting that $\Gamma$ changes only
due to rescaling. The analogue of Eq.(\ref{scalscale}) for a vector order
parameter is
$L(t)\sim (\rho_s\Gamma t/M^2)^{1/z}.$
Here $\rho_s$ is the large scale spin stiffness and $\rho\sim \zeta^{-(d-4)}$ in
the critical region. Thus we find, for a vector order parameter
\begin{equation}
L(t)\sim \zeta (t/\zeta^{z_c})^{1/4},
\label{scalevec}
\end{equation}
which is again consistent with the conventional dynamical scaling.
So we see $z=4$ for both $N=2$ and $N=3$. The $N=4$ case is interesting due to
an extra logarithm: $L(t)\sim t^{1/4}[1+O(1/\ln t)]$, similar to the $N=2$ case of
non-conserved Landau-Ginzburg theory. However, RG cannot detect that.

It is easy to extend the large-$N$ method to the $(d-m)$ Lifshitz point:
The equation of motion is
\begin{equation}
{\partial \phi(k,t)\over\partial t}= 
\nabla_{\mid\mid} ^2\phi- \nabla^4 _{\perp}\phi +a(t)\phi,
\end{equation}
$a(t)$ being the same as in Eq.(\ref{nclarge}). 
Neglecting $db/dt$ (like the previous case), we get
\begin{equation}
1=\exp(2b)\int d^{d-m} k_{\mid\mid}  d^m k_{\perp}\exp(-2k_{\mid\mid}^2t)
\exp(-2k_{\perp}^4 t),
\end{equation}
giving $b= [(d-m)/4+m/8]\ln t$.
Hence the structure factor scales as
\begin{equation}
S(k,t)=t^{(d-m)/2}t^{m/4} \exp (-2k_{\mid\mid}^2 t-2k_{\perp}^4 t).
\label{ncopfinal}
\end{equation}
We see that there are two {\em dynamic exponents} $z_{\mid\mid}=1/2$ and
$z_{\perp}=1/4$ for the parallel and the perpendicular directions respectively, which
follows easily from the form of the bare propagator
for the equation of motion for the $(d-m)$ Lifshitz point is
$G(k,t)=\exp(-k_{\parallel}^2 t-k_{\perp}^4 t).$
This can be written in a scaling form 
$g(k_{\parallel}t^{1/2},
 k_{\perp}t^{1/4})$. The existence of two {\em dynamic exponents} is clear
from this form. The nonlinear terms in the long time limit only modify
the scaling form keeping the dynamic exponents same (within this 
self-consistent calculation). The results above immediately yield surprising results for
$\lambda$: As before we find
\begin{equation}
\phi_{\bf k}(t)=\phi_{\bf k}(0)t^{(d-m)/4+m/8}\exp[-k_{\parallel}^2t-k_{\perp}^4t].
\end{equation}
Thus the two-time correlator scales as
\begin{eqnarray}
C(k,t,t')&=&t^{(d-m)/2+m/4}\left({t'\over t}\right)^{(d-m)/4}\left({t'\over
t}\right)^{m/8}\nonumber \\ &&\times
\exp[-k_{\parallel}^2(t+t')-k_{\perp}^4(t+t')].
\end{eqnarray}
Thus we identify two $\lambda's$: $\lambda_{\parallel}=(d-m)/2$ and
$\lambda_{\perp}=m/2$. So in presence of anisotropy, like the dynamic 
exponents, one finds two two-time correlation exponents. Again it is
important to find out how $O(1/N)$ corrections may affect the effects of
anisotropy on the $\lambda$-exponents.

Folk and Selke \cite{folk} studied the relaxation near the Lifshitz point. They
characterised relaxation rate by two exponents: a single dynamic exponent $z$ and
a second exponent $x$ which characterises spatial anisotropy. Our
Eq.(\ref{ncopfinal}) can also be written in terms a single dynamic exponent $z$
and a new exponent $x$:
\begin{eqnarray}
S(k,t)&=&t^{(d-m)/2+m/4}\exp[-2k_{\parallel}^2t-2k_{perp}^4t]\nonumber \\
&=&t^{(d-m)/z+m/(xz)}\exp[-2k_{\parallel}^zt (1+k_{perp}^x/k_{\parallel})]
\end{eqnarray}
with $z=2$ and $x=2$. Thus our results agree with the scaling form at the Lifshitz
point.

\subsection{Conserved order parameter}
Having considered the simpler case of quench of the nonconserved order
parameter, we now go over to the more interesting case of conserved
order parameter. We first work out the $(d-d)$ Lifshitz point using the
large-$N$ method: In the large-$N$ limit the equation of motion 
in Fourier space becomes  \cite{bray}
\begin{equation}
{\partial \phi\over\partial t}=-k^2[Dk^4 +a(t)]\phi.
\end{equation}
The solution of this is
\begin{equation}
\phi(k,t)=\phi(k,0)\exp[-k^2b(t)-k^6t].
\end{equation}
As usual, $b(t)=\int_0^t a(t')dt'$. We solve for $b$
self-consistently. Using the solution of the equation of motion we have
$a(t)={db\over\ dt}=1-\Delta\int_k d^dk \exp[2k^2b-2k^6t].$
Neglecting $a(t)$ in comparison with 1 (which we jusitify {\em aposteriori}),
\begin{eqnarray}
1= \Delta (b/t)^{d/4}\int dy y^{d-1} \exp [2\lambda (y^2-y^6)],
\end{eqnarray}
with ${\bf k}=(b/t)^{1/4}{\bf y}$ and $\lambda=b^{3/2}/\sqrt{t}$. 
Evaluating the above integral by using steepest descent method we obtain
$b= Ct^{1/3}(\ln t)^{2/3}$,
($C$ is a constant) in the long time limit. So 
the asymptotic form of the structure factor is
\begin{equation}
S(k,t)\approx t^{{d/6}\theta(k/k_m)}, k_m=({d\ln t\over\ t})^{1/6},
\end{equation}
where $\theta (x)=3(a^{2/3}x^2-x^6)$. Now 
$a(t)=db/dt=-(C/3)\,t^{-2/3}(\ln t)^{2/3}-(C/3)\,t^{-2/3} 1/(\ln t)^{-1/3} 
\rightarrow 0$ as $t\rightarrow \infty$.
Thus there are two time scales present in $S(k,t)$, separated by
$\ln t$; in other words this form, instead of simple scaling,
exhibits multiscaling similar to the quench of
conserved ordinary Ising order parameter \cite{bray,cog}. 
Following \cite{bray} it is easy to show that
the equal time correlation function $C(x,t)\equiv \langle {\bf \phi}({\bf x},t)
\cdot {\bf \phi}(0,t)\rangle$ satisfy the equation
\begin{equation}
{1\over 2}{\partial C\over\partial t}=\nabla^6 C- a(t)\nabla^2(C+{C^3\over N}).
\label{finn}
\end{equation}
Equation (\ref{finn}) is similar to that in Ref.\cite{hum}.
This however is not correct to $O(1/N)$ as pointed out in Ref.\cite{cas1}. 
This can be derived by using the Gaussian auxiliary field
method of Mazenko \cite{maz}.
For $N=\infty$, the $C^3/N$ term can be ignored and the
resulting equation exhibits temporal multiscaling.
For finite $N$ we start with a scaling form consistent with $t^{1/6}$ growth
law, i.e., $S(k,t)=t^{d/6}g(kt^{d/6}),\,a(t)=q_m^2/t^{2/3}$, and
$C(r,t)=f(r/t^{1/6})$, where $f(x)$ is the Fourier transform of $g(q)$ and
$q_m$ is a constant. Substituting these in Eq.(\ref{finn}) gives
\begin{equation}
{dg\over dq}=-\left({d\over q}+12q^5-12q_m^2q\right),\;\;B(q)={8q_m^2\over N}
(f^3)_{\bf q},
\end{equation}
where $(f^3)_{\bf q}$ is the Fourier transform of $f(x)^3$. We have $g(0)=0$
for a conserve order parameter. We have the constraint $f(0)=1$, i.e., $\sum_q g(q)=1$
to fix $q_m$ \cite{bray}. 
For large $N$ we find that 
$q_m\sim (\ln N)^{1/3}$ which diverges with $N$. 
This defines a lengthscale $L(t)\sim
t^{1/6}$. Thus, as in the Ising case, our approximate analysis gives 
consistent scaling solutions. So we do not find any
temporal multiscaling for finite (but large) $N$. Fourier transform of $g(q)$
gives the scaling function $f(x)$. Recently, in a series of papers Castellano
{\em et al} \cite{cas1,cas2,cas3} discussed 
preasymptotic multiscaling and eventual crossover to simple scaling
in the asymptotic limit for the COP of the Ising model. They showed that 
there are competitions between terms representing multiscaling
and simple scaling respectively; and consequently they found a crossover time
scale beyond which simple scaling is recovered. Analogous analysis on
Eq.(\ref{finn}) reveals similar features.

In a RG set up for COP, 
$\Gamma$ should be replaced by $\Gamma k^2$ in the
equation of motion that we used while discussing the RG arguments for the
nonconserved order parameter.
After suitable coarse graining and rescaling of variables, the equation of
motion reads (under rescaling $k'=k/b,t'=b^zt$)
\begin{equation}
b^{2-z+2\chi-y}(1/\Gamma {k'}^2) {\partial \phi' _{k'}\over\partial t'}
=-{\delta F\over\delta \phi' _{-k'}}+\zeta'_{k'}(t')/(\Gamma k'^2)
\end{equation}
while the new noise correlator is given by
\begin{equation}
<\zeta'_{k'}(t_1')\zeta'_{k'}(t_2')>=b^{2-z+2\chi-2y}2T\Gamma k'^2
\delta(t_1'-t_2')
\end{equation}
As usual the temperature scales as $T'=b^{-y}T$.
Since the noise correlator is singular in the $k\rightarrow 0$ limit,
no new term can be generated by integrating out the short wavelength
fluctuations which is at least as singular as the bare correlator \cite{bray}. 
Hence in
the long wavelength limit, the bare correlator dominates and consequently
we claim that $1/\Gamma$ can change only under rescaling. 
We write the recursion relation for $1/\Gamma$:
\begin{equation}
(1/\Gamma')=b^{2-z+2\chi-y}(1/\Gamma).
\end{equation}
Provided $1/\Gamma$ is nonzero at the fixed point, we have
$z=2+2\chi-y=d+2-y$
where we used $2\chi=d$. As before we have (from Porod law; see previous Sec.)
$y=d-1$ for $N=1$, $y=d-2$ for
$N=2$, $y=d-3$ for $N=3$ and
$y=d-4$ for $N\geq 4$. Plugging in these values we 
get $z=3$ for
scalar order parameter ({\em Lifshitz-Slyzov} law for this model), 
$z=4$ for $N=2$ component case, $z=5$ for $N=3$ and 
$z=6$ for the large-$N$ case. For $N=4$ there is additional logarithm
in the form: $L(t)\sim t^{1/6}[1+O(1/\ln t)]$; our RG treatment cannot see this extra
logarithm. It is surprising that we find $z$ for the ordering of 
non-conserved dynamics of the scalar order parameter ($N=1$) higher than that
in the conserved dymamics, as normally conservation
law is expected to slow down the dynamics. Since RG for NCOP is not exact it is
difficult to interpret this surprising result. Numerical simulations are
required to settle this.

We now show a simple argument leads to the $t^{1/4}$ growth for the $N=1$ 
scalar conserve order
parameter (COP). The diffusion current ${\bf J}\sim -\nabla\mu$ with $\mu$ being the
chemical potential difference between the two species. If there is only one scale $L(t)$
at large times then $\nabla \mu\sim \Delta\mu /L$ where $\Delta \mu$ is the change in
$\mu$. Near coexistence, the free enery difference is of order $\phi\Delta\mu$ (where
$\phi$ is the magnitude of the order parameter in either of the phases).
Since in the nucleation of NCOP, there is a critical length scale set by $\sigma$ the
surface-energy and the free energy difference 
$\Delta f$: $L\sim \sigma/\Delta, f
\sim \sigma/\phi\Delta \mu$. Thus we get $J\sim\sigma\phi L^2$. Since $J\sim\phi
\dot{L}$ we have 
\begin{equation}
{\partial L\over\partial t}\sim {1\over L^2}.
\end{equation}
This gives $L(t)\sim t^{1/3}$ at long times, which is the analogue of the {\em
Lifshitz-Slyzov} law for this model. This prediction is in accordance with our RG result
above.

We extend a simple argument
due to J. Das and M. Rao \cite{jay} to study the effects of the mode-coupling terms on
the ordering dynamics. In presence of a 
mode-coupling term, the Eq. of motion 
with the free energy (\ref{free}) is
\begin{equation}
{\partial \phi_{\alpha}\over\partial t}=\Gamma\nabla^2{\delta F\over\delta 
\phi_{\alpha}}+g\epsilon_{\alpha\beta\gamma}\phi_{\beta}\nabla^4\phi_{\gamma}
+\eta_{\alpha}.
\label{mode}
\end{equation}
Notice that the precisional term involves $\nabla^4$ instead of $\nabla^2$ as
in the precisional dynamics of Heisenberg ferromagnet. We write Eq.(\ref{mode})
as  (in the spirit of dimensional analysis)
a continuity equation and replace the corresponding current 
 by the `velocity' ${dL\over dt}$ to obtain
\begin{equation}
{dL\over dt}=\Gamma{\sigma\over L^4}+g{\sigma M_o\over L^3},
\end{equation}
where $M_o,\sigma$ and $\Gamma^{-1}$ are the equilibrium magnetisation, surface tension and spin mobility respectively. It is easy to see that beyond a
crossover time given by $t_C(G)\sim (\Gamma/M_o g)^5\sim 1/G^5$ dynamics
crosses over from $z=5$ to $z=4$. Thus we expect that in presence of a 
torque dynamic exponent is 4, showing the relevance of the torque.

One can easily extend the large-$N$ calculation to the $(d-m)$ Lifshitz point:
The equation of motion is
\begin{eqnarray}
{\partial \phi\over\partial t}=-(k_{\perp}^2+k_{\parallel}^2)[k_{\parallel}^2
+k_{\perp}^4 +a(t)]\phi.
\label{dml}
\end{eqnarray}
In the long wavelength limit the solution of Eq.(\ref{dml}) is (retaining terms
up to $O(k^4)$) 
\begin{equation}
\phi(k,t)=\phi(k,0)\exp[k^2b(t)-k_{\parallel}^4t-k_{\parallel}^2
k_{\perp}^2t],
\end{equation}
with $b(t)=\int_0^t a(t')dt'$. Neglecting $a(t)$ we obtain
$1 =(b/t)^{d/2}\int d^m x_{\parallel}d^{d-m}x_{\perp}\exp[2\beta 
(x_{\parallel}^2-x_{\parallel}^2x_{\perp}^2-x_{\parallel}^4)].$
Here ${\bf k=x}(b/t)^{1/2},\beta=b^2/t$. By using steepest decent method we
obtain 
\begin{equation}
S(k,t)=t^{{(d/4)}\theta'(k/k_m)}; k_m=({d\ln t\over\ t})^{1/4},
\end{equation}
where $\theta' (k/k_m)$ another function of $k/k_m$. It is easy to see
that $a(t)$ indeed goes to 0 as $t\rightarrow \infty$ justifying our
assumption. 
Here also we find multiscaling - there are two time scales separated by
$\ln t$. Presumably, this is also an artefact of the $N\rightarrow \infty$
limit as in the $(d-d)$ Lifshitz point case. 
Interestingly, however we do not find separate timescales for
$\parallel$ and $\perp$ directions unlike the nonconserved case.
The signature of this is already there in the bare propagator:
$G(k,t)=\exp[k^2(k_{\parallel}^2+k_{\perp}^4)t]=
\exp[(k_{\parallel}^2k_{\perp}^2+k_{\parallel}^4)t]$
in the limit $k_{\perp}\rightarrow 0,
k_{\parallel}\rightarrow 0$.
Thus under the scaling ${\bf k}\rightarrow {\bf k}/b,
t \rightarrow b^4 t, G$ does not change implying $z=4$. Here also
nonlinearities determine the scaling form keeping $z=4$. It is interesting
to note that for the case of $(d-m)$ Lifshitz point for NCOP
we obtain two distinct $z$'s reflecting the anisotropy. However for COP
the anisotropy does not show up, i.e., one obtains only one $z=4$ in the
lowest order. The reason for this can be easily understood from the 
above analysis. 

\section{Scaling in a temperature quench from the paramagnetic to the 
modulated phase}
In the above we have so far considered temperature quench through the 
Lifshitz point
from the paramagnetic to the ferromagnetic phase. In these cases 
the initial states
are random (no order) and the final ferromagnetic states are uniform (i.e., the
ordering wavevector is 0). However, if the system is quenched from high- to
low- temperatures through an off-Lifshitz point path there are two more cases
possible: i)From the paramagnetic to ferromagnetic (along the line AB in
Fig.(\ref{fig1})): In this case the ordering
dynamics in the long wavelength limit 
is same as that of an Ising order parameter which has been studied
extensively; and ii)from the paramagnetic to the modulated phase (along the path
EF in Fig.(\ref{fig1})). We examine the growth of order in the
second case in a large-N set up. For simplicity we consider the dynamics of
the non-conserved order parameter (NCOP) only.

To calculate the form of the structure factor after a quench from the
paramagnetic phase to the modulated phase for the NCOP we start with a free
energy $F=\int d^dx [-{1\over 2}\phi^2 +
{1\over 2}(\nabla_{\mid\mid}\phi)^2-{c_{\perp}\over 2}
(\nabla_{\perp}\phi)^2+{D\over 2}(\nabla^2\phi)^2+u(\phi^2)^2],\;\phi^2
=\phi_{\nu}\phi_{\nu},\nu=1,...,N$ for the modulated phase we obtain the
equation of motion [we consider the general case of a system with ($d-m$)
Lifshitz point]
\begin{eqnarray}
&&{\partial\phi\over\partial t}=-{\delta F\over\delta \phi}\nonumber \\
&=&\phi+\nabla_{\parallel}^2\phi-c_{\perp}\nabla_{\perp}^2\phi
-D\nabla_{\perp}^4 \phi -{u\over N}\phi (\phi_{\nu}\phi_{\nu})\nonumber\\
&=&\nabla_{\parallel}^2\phi-c_{\perp}\nabla_{\perp}^2\phi
-D\nabla_{\perp}^4 \phi+a(t)\phi.
\label{moduleq}
\end{eqnarray}
Equation (\ref{moduleq}), as before, is exact in the $N\rightarrow \infty$
limit. Here, $a(t)=1-\langle\phi^2\rangle$. The intial condition is 
\begin{equation}
\langle \phi_{\bf k}(0)\phi_{-\bf k}(0)\rangle=\Delta.
\label{inimod}
\end{equation}
The solution of Eq.(\ref{moduleq}) is
\begin{equation}
\phi_{\bf k}(t)=\phi_{\bf k}(0)\exp[-k_{\parallel}^2t+c_{\perp}k_{\perp}^2 t-
Dk_{\perp}^4t+b(t)],
\label{modsol}
\end{equation}
with $a(t)={db\over dt}$. By using the initial condition Eq.(\ref{inimod}) we
find for the equal time correlation function or the structure factor
\begin{eqnarray}
S({\bf k},t)&\equiv&\langle\phi_{\bf k}(t)\phi_{\bf -k}(t)\rangle\nonumber \\
&=& \Delta
\exp [-2k_{\parallel}^2t+2c_{\perp}k_{\perp}^2 t-2Dk_{\perp}^4t+2b(t)].
\end{eqnarray}
Here $b(t)$ has to be evaluated self-consistently. As before neglecting $a(t)$
in the long $t$ limit (which we justify {\em a posteriori}) we find
\begin{eqnarray}
1&=&\Delta\int d^dk\exp[-2k_{\parallel}^2t+2c_{\perp}k_{\perp}^2
t-2Dk_{\perp}^4t+2b(t)]\nonumber \\
&=&\Delta\exp(2b(t))\int d^{d-m}k_{\parallel}d^mk_{\perp}
\exp(-2k_{\parallel}^2t)\nonumber \\
&&\times\exp(2c_{\perp}k_{\perp}^2 t-2Dk_{\perp}^4t).
\end{eqnarray}
The integration over $k_{\parallel}$ can be done trivially. For $k_{\perp}$ we
use a saddle point approximation in the long time limit: We expand $F(k_{\perp})
=c_{\perp}k_{\perp}^2 -Dk_{\perp}^4$ about $k_{\perp}= 
\sqrt {c_{\perp}\over 2D}\equiv q_o$ where $q_o$ is the ordering wavevector in
the modulated phase. In the long time limit we find
\begin{equation}
b(t)=F(q_o)\ln [t^{d-m\over 4}t^{1/4}].
\label{bmod}
\end{equation}
This gives $S({\bf k},t)$ as 
\begin{eqnarray}
S({\bf k},t)&\sim& t^{d-m+1\over 2}\times \nonumber \\
&&\exp [-2k_{\parallel}^2t 
-2(q_o^2-c_{\perp}k_{\perp}^2)t-2(Dk_{\perp}^4 -q_o^4)t].
\label{sformod}
\end{eqnarray}
Equation (\ref{sformod}) clearly violates the standard form 
Eq.(\ref{stanform}) for the structure factor as there is a distinct lenght scale
$q_0^{-1}=\sqrt {2D/c_{\perp}}$. We can identify a dynamic
exponent $z_{\parallel}=2$ for the parallel direction by comparing with
Eq.(\ref{stanform}). However for the perpendicular directions dynamic scaling
of the standard form is not found. This feature manifests more clearly for the
($d-d$) Lifstitz point case where the structure factor in the long time limit
assumes the form
\begin{equation}
S({\bf k},t)\sim t^{1/2}\exp [-2(q_o^2-c_{\perp}k^2)t-2(Dk^4 -q_o^4)t],
\end{equation}
which clearly does not obey the standard dynamic scaling form. It is not
apriori clear if this is an artefact of the large-N method employed here or
whether standard dynamic scaling is recovered when corrections to $O(1/N)$ are
included. This requires further investigations.
It is however clear from Eq.(\ref{sformod}) that in the long time limit
along the parallel directions
$S({\bf k},t)$ is non-zero for $k_{\parallel}=0$ and along the perpendicular
directions $S({\bf k},t)$ is non-zero along $k_{\perp}=q_o$. Thus 
$k_{\parallel}=0$ and 
$k_{\perp}=q_o$ are the ordering wavevectors, as expected. It is also clear
from Eq.(\ref{bmod}) that in the large $t$ limit, $a(t)\rightarrow 0$
justifying our assumption.

We obtain, from Eq.(\ref{modsol}) the two-time correlation function for the
($d-m$) case as
\begin{eqnarray}
C({\bf k},t,t')&\sim& t^{d-m+1\over 2} \left({t'\over t}\right)^{(d-m)/4}
\left({t'\over t}\right)^{1/4}\times\nonumber \\
\exp [-k_{\parallel}^2(t+t')&&-
(q_o^2-k^2)(t+t')-(Dk^4 -q_o^4)(t+t')].
\end{eqnarray}
Thus comparing with the standard form Eq.(\ref{stanc}) 
we can identify a two-time 
exponent for the parallel directions: $\lambda_{\parallel}=(d-m)/2$. We also
detect the existence of a new exponent $\tilde{\lambda}=1/4$ coming from the
perpendicular directions which is {\em not} of the standard form. Thus, in
the $N\rightarrow \infty$ limit, after a quench from the paramagnetic to the
modulated phase the standard form of the correlation function does not hold.

Kaski {\em et al} studied the scaling properties of the anisotropic structure
factor after a quench from the paramagnetic to the modulated phase \cite{gunton}
in a two dimensional ANNNI model ($d=2,m=1$ in our notation) employing Glauber
spin dynamics (i.e., corresponding to the NCOP case studied here). There principal
findings include (i)a dynamic exponent $z=2$ for all directions and 
(ii)anisotropic growth rates. These results agree qualitatively  with our 
large-$N$ results given by Eq.(\ref{sformod}) with $d=2,m=1$ for which the
structure factor is given by
\begin{equation}
S({\bf k},t)=t\exp[-2k_{\parallel}^2t
-2(q_o^2-c_{\perp}k_{\perp}^2)t-2(Dk_{\perp}^4 -q_o^4)t].
\end{equation}
Thus $S(0,t)$ grows as $t$ yielding (in the notation of \cite{gunton}) $n\approx
0.5$ and for finite wavevectors this growth is modulated anisotropically, as
discussed in Ref.\cite{gunton}.

\section{Conclusion}
We analyse the form of the correlation and structure factors after a quench
in a system with a Lifshitz point. We first consider a quench through the
Lifshitz point from the
paramagnetic to the ferromagnetic phase. We studied both the nonconserved and 
conserved
order parameters. In particular we obtained the dynamic exponent $z$ by
using  large-$N$ as well as RG calculations. We also calculate the two-time
exponent $\lambda$ for the nonconserved order parameter. We find multiscaling 
for the conserved case in the $N\rightarrow \infty$ limit 
and simple scaling for the nonconserved case in our one-loop calculations. 
We however show that this multiscaling is an artefact of the $N\rightarrow
\infty$ limit and for finite (but large) $N$ simple scaling is recovered. We construct
Allen-Cahn and Lifshitz-Slyzov type scaling laws for the nonconserved and conserved
order parameters respectively.
We have also shown that in the anisotropic $(d-m)$ 
case, for a conserved order parmeter
anisotropy is unimportant (irrelevant in a RG sense)
where as for a non-conserved order parameter it shows
up in the form of two dynamic and two two-time exponents. 
It is important to see if anisotropy remains unimportant in higher order
approximations and/or numerical simulations.  We argue that
in presence of a torque in the Eq. of motion for a conserved order parameter
for the $(d-d)$ Lifshitz point $z$ reduces from 5 to 4. We have also
discussed RG arguments in the context of our one-loop results. 
We also show, in the $N\rightarrow\infty$ limit that the standard form of the
structure factor exhibiting dynamical scaling breaks down 
after a quench from the
paramagnetic phase to the modulated phase. It is not clear if the breakdown of
dynamical scaling will persist even after the inclusion of corrections 
of $O(1/N)$. This remains an open problem. 
Our results can be easily checked numerically by extending the cell dynamics
methods \cite{puri} as appropriate for systems with Lifshitz points. 
Numerical measurements of higher-order correlation
functions will help us to understand the validity of the Mazenko method in this
problem. Recent numerical studies \cite{jay} suggest that the Mazenco
method may not be very good for COP for Ising systems. 
Numerical simulations are required to
settle this for systems with Lifshitz points. In short, we have analysed the
asymptotic form of the correlation functions/structure factors after
temperature quenches in systems with Lifshitz points. We feel our results
are sufficiently important and interesting 
to induce further detailed studies of these
phenomena in systems with Lifshitz points.

\section{Acknowledgement}
One of the authors (AB) wishes to thank Jayajit Das for helpful discussions. AB
thanks the Alexander von Humboldt Foundation, Germany for financial support.

\end{multicols}
\end{document}